\documentclass[aps,pre,showpacs,twocolumn]{revtex4}%
\usepackage[english]{babel}
\usepackage{amsmath}
\usepackage{epsfig}
\usepackage{graphicx}
\usepackage{dsfont} 
\usepackage[usenames,dvipsnames]{color}

\usepackage{bm}
\usepackage{marvosym}
\usepackage{times,amsmath,amssymb}
\def\scs{\scriptstyle}

\newcommand{\be}{\begin{equation}}
\newcommand{\ee}{\end{equation}}
\newcommand{\bel}[1]{\begin{equation}\label{#1}}
\newcommand{\bea}{\begin{eqnarray}}
\newcommand{\eea}{\end{eqnarray}}
\newcommand{\balign}{\begin{align}}
\newcommand{\ealign}{\end{align}}
\newcommand{\ba}{\begin{array}}
\newcommand{\ea}{\end{array}}
\newcommand{\bfig}{\begin{figure}}
\newcommand{\efig}{\end{figure}}

\newcommand{\eref}[1]{(\ref{#1})}

\newcommand{\rmd}{\mathrm{d}}
\newcommand{\rme}{\mathrm{e}}

\newcommand{\eps}{\varepsilon}

\newcommand{\bra}[1]{\mbox{$\langle \, {#1}\, |$}}
\newcommand{\ket}[1]{\mbox{$| \, {#1}\, \rangle$}}

\newcommand{\comm}[2]{\mbox{$[\,{#1}\,,\,{#2}\,]$}}
\newcommand{\BIN}[2]{\renewcommand{\arraystretch}{0.8}\mbox{$\left(\ba{@{}c@{}}{\scs #1}\\{\scs #2}\ea\right)$}
\renewcommand{\arraystretch}{1}}
\newcommand{\Tr}{\mathop{\mathrm{Tr}}\nolimits}

\newcommand{\C}{{\mathbb C}}

\newcommand{\N}{{\mathbb N}}

\begin{document}
\title{Exact matrix product solution for the boundary-driven Lindblad $XXZ$-chain}
\author{D. Karevski$^{1}$, V. Popkov$^{2,3}$, and G.M. Sch\"utz$^{4}$}
\affiliation{$^{1}$ Institut Jean Lamour, dpt. P2M, Groupe de
Physique Statistique, Universit\'e de Lorraine, CNRS, B.P. 70239,
F-54506 Vandoeuvre les Nancy Cedex, France} \affiliation{$^{2}$
Dipartimento di Fisica, Universit\`a di Firenze, via Sansone 1,
50019 Sesto Fiorentino (FI), Italy} \affiliation{$^{3}$ Max Planck
Institute for Complex Systems, N\"othnitzer Stra{\ss }e 38, 01187
Dresden, Germany} \affiliation{$^{4}$Institute of Complex Systems
II, Forschungszentrum J\"ulich, 52428 J\"ulich, Germany}
\begin{abstract}
We demonstrate that the exact non-equilibrium steady state of the
one-dimensional Heisenberg XXZ spin chain driven by boundary
Lindblad operators can be constructed explicitly with a matrix
product ansatz for the non-equilibrium density matrix where the
matrices satisfy a {\it quadratic algebra}. This algebra turns out
to be related to the quantum algebra $U_q[SU(2)]$. Coherent state
techniques are introduced for the exact solution of the isotropic
Heisenberg chain with and without quantum boundary fields and
Lindblad terms that correspond to two different completely
polarized boundary states.
We show that
this boundary twist leads to non-vanishing stationary currents of
all spin components. Our results suggest that the matrix product
ansatz can be extended to more general quantum systems kept far
from equilibrium by Lindblad boundary terms.
\end{abstract}

\date{\today }
\maketitle

The non-equilibrium behaviour of open quantum systems has become
accessible through recent advances in artificially assembled
nanomagnets consisting of just a few  atoms \cite{ScienceLoth11}
or in the study of quasi one-dimensional spin chain materials like
SrCuO$_{2}$ where many transport characteristics are measurable
experimentally \cite{spinchain,HessBallistic2010}.
In particular, it is desirable to understand
the interplay between many-body \textit{bulk} properties (e.g. magnon excitations or
magnetization currents in quantum spin systems) and
\textit{local} pumping (applied to the boundary of a system) driving the
system constantly out of equilibrium. A good starting point is provided by
the anisotropic Heisenberg model
\cite{Baxt82}
\bel{1-1}
H =  J \sum_k (\sigma_k^x  \sigma_{k+1}^x + \sigma_k^y  \sigma_{k+1}^y
+ \Delta \sigma_k^z  \sigma_{k+1}^z - \eps_0)
\ee
of coupled spins. The pure quantum version of this model is
exactly solvable by the Bethe ansatz.
Interestingly, within linear response theory, i.e., close to equilibrium, it was found that
at finite temperature a diffusive contribution to the Drude weight appears
\cite{Sirk11,Jese11,Stei12}, which is at
variance with the long-held belief that integrability protects the ballistic nature of transport
phenomena. Unfortunately the Bethe-ansatz fails
in the more relevant context of open far-from-equilibrium systems
where these questions can be addressed directly
in terms of the Lindblad Master equation \cite{Petruccione}
\bel{1-2}
\frac{\rmd}{\rmd t} \rho  = - i \left[  H,\rho \right]  + \mathcal{D}^L(\rho) + \mathcal{D}^R(\rho)
\ee
for the reduced density matrix $\rho$ associated to the chain
(here and below we set $\hbar=1$).
The dissipative terms $\mathcal{D}^{L,R}(\rho)
= D^{L,R}\rho {D^{L,R}}^{\dagger} -  \frac{1}{2}\left\{  \rho, D^{L,R} {D^{L,R}}^{\dagger} \right\}$
with the Lindblad operators $D^{L,R}$ acting
locally at the open ends of the
quantum chain (see below) describe the coupling to external reservoirs that drive a current
through the system and thus keep the
system in a permanent non-equilibrium steady state.
Indeed, using dissipative dynamics for the preparation of quantum
states is becoming a promising field of
research \cite{ZollerNature2008,ZollerPRA}.

Significant progress has been achieved very recently in two
remarkable papers by Prosen \cite{Pros11a,Pros11b} who observed
that the exact stationary density matrix for the XXZ chain with
one specific pair of Lindblad boundary terms can be constructed
explicitly in matrix product operator form \cite{Vers04}
by a matrix product ansatz (MPA) somewhat reminiscent of the
matrix product ansatz of
Derrida et al. \cite{Derr93} for the
stationary distribution of purely classical stochastic dynamics.
With an explicit
representation of the matrix algebra Prosen was then able to
compute analytically various physical quantities of interest.
However, in contrast to \cite{Derr93}, where the matrices
satisfy a quadratic algebra, the matrices of
\cite{Pros11a,Pros11b} 
satisfy a cubic algebra which arises from a peculiar local cancellation
mechanism involving three neighboring sites in the quantum chain.
This feature is significant since, due to the lack of a general representation
theory for cubic algebras,
this approach does not lend itself easily to generalization to other
open quantum systems with other cubic algebras
or even small modifications of the original problem
such as boundary fields or other Lindblad terms for the XXZ chain
which would require a different representation.
Indeed, the wide applicability of the MPA of \cite{Derr93}
derives from the fact that many quadratic algebras (which include
all Lie algebras through their commutation relations) have explicitly known
representations which is crucial for the exact computation
of physical observables \cite{Blyt07}.

In this Letter we show that exact non-equilibrium steady states for open quantum systems
can be obtained from a matrix product ansatz
which yields a {\it quadratic} algebra.  Specifically, we consider the Lindblad quantum XXZ chain
and show
that the associated matrix algebra is related to the bulk symmetry of the XXZ-chain,
which is the quantum algebra $U_q[SU(2)]$ with $\Delta = (q+q^{-1})/2$.
A coherent state representation makes it possible to consider
Lindblad operators that correspond to two different
completely polarized boundary states, viz., in the $(y,z)$ plane on the left boundary
\bel{1-3}
D^L = \sqrt{\frac{\Gamma}{2}} (\sigma^x_1 + i \cos{\theta_L} \sigma^y_1 - i \sin{\theta_L} \sigma^z_1 )
\ee
and in the $(x,z)$ plane on the right boundary  with
\bel{1-3a}
D^R = \sqrt{\frac{\Gamma}{2}}  ( \cos{\theta_R} \sigma^x_N - i \sigma^y_N + \sin{\theta_R} \sigma^z_N ).
\ee
These Lindblad generators
lead to local dissipative terms
whose stationary solutions, satisfying ${\cal D}^{L(R)}(\rho^{L(R)})=0$, are respectively
the pure states $\rho^L=1/2(\mathds{1}+\sigma^z_u)=\ket{\uparrow_u}\bra{\uparrow_u}$
and $\rho^R=1/2(\mathds{1}-\sigma^z_v)=\ket{\downarrow_v}\bra{\downarrow_v}$,
where $\ket{\uparrow_u}$ is the eigenstate associated to the eigenvalue $+1$ of
$\sigma^z_u=\sin\theta_L \sigma^y+\cos\theta_L \sigma^z$, and
$\ket{\downarrow_v}$ is the eigenstate of
$\sigma^z_u=-\sin\theta_R \sigma^x+\cos\theta_R \sigma^z$ with eigenvalue $-1$.
For computational convenience we have chosen equal left and right amplitudes $\Gamma$
in \eref{1-3} and \eref{1-3a}. By a judiciously chosen similarity transformation these amplitudes
can be made different \cite{Pros12}.

Moreover, we allow for quantum boundary fields acting on the directions of the local polarizations
specified by \eref{1-3} and \eref{1-3a}. Consequently  we add to the Hamiltonian \eref{1-1} the
contribution $\vec{f}^L  \cdot \vec{\sigma}=f^L \sigma^z_u$ for the left-end of the chain  and
$\vec{f}^R  \cdot \vec{\sigma}=f^R \sigma^z_v$ coming from the right-end boundary field.
For convenience we choose $J=1/2$ and $\eps = 1$
so that $H =  \sum_{k=1}^{N-1} h_{k,k+1} + g^L_1 + g^R_N $
with the four-by-four matrix
$h = \frac{1}{2}(\sigma^x \otimes \sigma^x + \sigma^y \otimes \sigma^y + \Delta(\sigma^z \otimes \sigma^z -1)$
for the nearest neighbour bulk interaction and the two-by-two matrices
$g^{L} = f^{L} \sigma^z_u$ and $g^{R} = f^{R} \sigma^z_v$ for the boundary fields.
The subscript indicates on which sites of the chain the quantum operators $g$ and
$h$ act non-trivially. We write the stationary density matrix satisfying
\bel{1-2a}
i \left[  H,\rho \right]  = \mathcal{D}^L(\rho) + \mathcal{D}^R(\rho)
\ee
in the standard form $\rho = SS^\dagger / \Tr(SS^\dagger)$.

Our starting point for solving \eref{1-2a} is a matrix product ansatz
\bel{1-4}
S = \bra{\phi} \Omega^{\otimes N} \ket{\psi}
\ee
which we augment by auxiliary
matrices $\Xi$ such that the {\it local divergence condition }
\bel{cancel}
\comm{h}{\Omega \otimes \Omega} = \Xi \otimes \Omega -  \Omega \otimes \Xi
\ee
is satisfied.
In this construction
\bel{OmegaXi}
\Omega = \left( \ba{ll} A_1 & A_+ \\ A_- & A_2 \ea \right), \quad
\Xi = \left( \ba{ll} E_1 & E_+ \\ E_- & E_2 \ea \right).
\ee
are two-by-two matrices
whose entries $A_a, E_a$ are non-commuting matrices that act in a space
$\mathcal{A}$ with inner product $\bra{\cdot}\,\cdot \rangle$, and $\bra{\phi},\, \ket{\psi}$ are vectors
in $\mathcal{A}$.
In terms of Pauli matrices $\sigma^\pm = \frac{1}{2}(\sigma^x\pm i\sigma^y)$, $\sigma^z$,
and the two-dimensional unit matrix $\mathds{1}$ one can conveniently write
$\Omega = A_0 \mathds{1} + \vec{A} \cdot \vec{\sigma}$ with $A_0=\frac{1}{2}(A_1 + A_2)$,
$A_z=\frac{1}{2}(A_1 - A_2)$, $A_x=\frac{1}{2}(A_+ + A_-)$, $A_y=\frac{i}{2}(A_+ - A_-)$.
In our construction
the local divergence condition leads to a set of 16 {\it quadratic} relations for the eight matrices
$A_a, E_a$ and the problem to be attacked is the construction of matrices which satisfy these
relations.

Remarkably, all 16 equations \eref{cancel} can be solved
in terms of only three independent matrices $A_\pm,\,Q$
with $Q Q^{-1} = Q^{-1} Q = 1$
by choosing the auxiliary matrices
$E_\pm=0$, $E_1 = (q-q^{-1})(bQ - cQ^{-1})/2$, $E_2 = - (q-q^{-1})(\bar{b}Q - \bar{c}Q^{-1})/2$,
setting
\bel{1-7}
A_1 = bQ + cQ^{-1}, \quad A_2 = \bar{b} Q + \bar{c} Q^{-1}
\ee
and requiring
\bea
\comm{A_+}{A_-} & = & - (q-q^{-1})(b\bar{b} Q^2 - c\bar{c} Q^{-2}) \\
\label{bulkalgebra}
Q A_\pm & = & q^{\pm 1} A_\pm Q.
\eea
The constants $b,c,\bar{b},\bar{c}$
are arbitrary. Choosing the parametrization
\bea
\label{1-8}
b =  \frac{\alpha}{q-q^{-1}} \frac{\nu}{\lambda} & & \bar{b} =  \frac{\alpha}{q-q^{-1}} \frac{1}{\lambda\nu} \\
\label{1-9}
c = - \frac{\alpha}{q-q^{-1}} \mu \lambda & & \bar{c} = - \frac{\alpha}{q-q^{-1}} \frac{\lambda}{\mu} ,
\eea
and defining
\bel{1-10}
A_\pm =: i \alpha S_\pm, \quad Q =: \lambda q^{S_z}
\ee
then leads to
\bea
\label{1-11}
\comm{S_+}{S_-} & = & \frac{q^{2S_z} - q^{-2S_z}}{q-q^{-1}} \\
\label{1-12} q^{S_z} S_\pm & = & q^{\pm 1} S_\pm q^{S_z} . \eea
These are the defining relations of $U_q[SU(2)]$, the $q$-deformed
universal enveloping algebra of the Lie algebra $SU(2)$, which is
the non-abelian symmetry of the bulk Hamiltonian \eref{1-1}
\cite{Pasq90}.

After deriving a matrix algebra from the bulk interactions the second step
is the explicit construction of such matrices and of the vectors $\bra{V}$ and $\ket{W}$
using the boundary interactions.
For the present case we note that
the representation theory of $U_q[SU(2)]$
is well-understood and analogous
to that of $SU(2)$, except when $q$ is a root of unity, where some special features arise
\cite{Pasq90}. In particular, with the definition
$[x]_q := \frac{q^x - q^{-x}}{q - q^{-1}}$ we have the
irreducible
representation (irrep)
\bea
S_z  & = & \sum_{k=0}^{\infty} (p-k) \ket{k}\bra{k} \nonumber \\
\label{1-13}
S_+ & = & \sum_{k=0}^{\infty} [k+1]_q \ket{k}\bra{k+1} \nonumber \\
S_- & = & \sum_{k=0}^{\infty} [2p-k]_q \ket{k+1}\bra{k} \nonumber
\eea
where $p$ is an arbitrary complex parameter. This irrep is infinite-dimensional, except when $2p \in \N$).
The bra's and ket's form an orthogonal basis of $\mathcal{A} = \C^\N$ with
inner product $\bra{k}\,k'\,\rangle = \delta_{k,k'}$. Relations \eref{1-7}, \eref{1-10}
then provide a representation of the matrices $A_a$.

In order to satisfy the boundary conditions involving the quantum boundary fields
and the Lindblad dissipators we further define $\Phi = \comm{g}{\Omega}$
and the tensor products
$\Xi_k = \Omega^{\otimes k-1} \otimes \Xi \otimes \Omega^{\otimes N-k}$,
$\Phi_1= \Phi \otimes \Omega^{\otimes N-1}$,
$\Phi_N = \Omega^{\otimes N-1} \otimes \Phi$.
The local divergence condition implies
$\comm{H}{\Omega^{\otimes N}} = \Phi^{L}_1 + \Xi_1  + \Phi^{R}_N  - \Xi_N$.
The stationary Lindblad equation  \eref{1-2a} can thus be split into two equations
\bea
\mathcal{D}^L(SS^\dagger) & = & i(\Phi^L_1 + \Xi_1)S^\dagger - i S({\Phi^L_1}^{\dagger} + \Xi_1^\dagger) \\
\mathcal{D}^R(SS^\dagger) & = & i(\Phi^R_N - \Xi_N)S^\dagger - i S({\Phi^R_N}^{\dagger} - \Xi_N^\dagger)
\label{stationary}
\eea
for each boundary.  Using the decomposition $S = \bra{\phi}[ \mathds{1}  A_0
+  \sigma^z A_z + A_+ \sigma^+ + A_- \sigma^-)] \otimes \Omega^{\otimes N-1}\ket{\psi}$ for the
first equation and $S = \bra{\phi} \Omega^{\otimes N-1} \otimes [ A_0 \mathds{1}
+ A_z  \sigma^z + A_+ \sigma^+ + A_- \sigma^-]\ket{\psi}$ for the second equation
and factoring out the term containing $\Omega^{\otimes N-1}$
yields two sets of equations for the action of the matrices
$A_a$ on the vectors $\bra{\phi}$ and $\ket{\psi}$ respectively.
In this letter we outline this programme
for the isotropic chain $\Delta=1$. The construction for $\Delta\neq 1$ is conceptually
similar, but technically more involved and will be presented in a more detailed paper
\cite{KPSlong}.

For taking the isotropic limit $q\to 1$ we choose the normalization factors $\alpha=\lambda=1$ and
set $\mu=\nu$ in \eref{1-8}, \eref{1-9}
and arrive at
\be
\Omega =
 \left( \ba{cc} \nu S_z &  i S_+ \\
i S_- &  \nu^{-1} S_z \ea \right), \quad
\Xi =
\left( \ba{cc}  \nu  & 0 \\
0 &  - \nu^{-1} \ea \right).
\ee
The quadratic relations for the quantum algebra turn into the usual commutation
relations $\comm{S_+}{S_-} = 2 S^z$, $\comm{S_z}{S_\pm} = \pm S_\pm$ for $SU(2)$.
The irreducible representation \eref{1-13} turns into an irrep of $SU(2)$ by observing
that $[x]_1 = x$. Since the Lindblad dissipators do not generate terms proportional to the
unit matrix, we cancel these terms that appear on the r.h.s. of \eref{1-2a} by setting
$\nu = i$ which leads to $A_0=0$ and $\Omega=i \vec{S} \cdot \vec{\sigma}$, where
$\vec{S}=(\frac{S_++S_-}{2},i\frac{S_+-S_-}{2},S_z)$ and
$\vec{\sigma}=(\sigma^x,\sigma^y,\sigma^z)$, or in terms of the $\sigma^{\pm}$ the form
$\Omega=i(S_z\sigma^z+ S_+ \sigma^+ +S_-\sigma^-)$.

The key step in solving the boundary equations is the introduction of coherent states
\bea
\label{1-16}
\bra{\phi} & := & \sum_{n=0}^\infty \frac{\phi^n}{n!} \bra{0} (S_+)^n = \sum_{n=0}^\infty \phi^n \bra{n} \\
\label{1-17}
\ket{\psi} & := & \sum_{n=0}^\infty \frac{\psi^n}{n!} (S_-)^n \ket{0}= \sum_{n=0}^\infty \psi^n \BIN{2p}{n} \ket{n}.
\eea
Using the commutation relations of $SU(2)$ one finds
\bel{1-18}
\bra{\phi} S_z = \bra{\phi}(p-\phi S_+), \quad \bra{\phi} S_- = \phi \bra{\phi}(2p-\phi S_+)
\ee
and
\bel{1-19}
S_z \ket{\psi} = (p-\psi S_-) \ket{\psi}, \quad S_+ \ket{\psi} = \psi (2p-\psi S_-) \ket{\psi}
\ee

The left boundary equations can now be solved by noting that the Lindblad operator \eref{1-3}
can be obtained from a complete polarization along the $z$-axis by the unitary transformation
$U=\rme^{i\frac{\theta_L}{2} \sigma^x}$ on site 1 of the chain which rotates the $z$-axis
into a new direction $u$. After this transformation the
leftmost matrix $\Omega$ in the tensor product $\Omega^{\otimes N}$ reads in the new basis
\be
\Omega(\theta_L)=  i\left(S_z(\theta_L)\sigma^z_u+
S_+(\theta_L) \sigma^+_u +S_-(\theta_L)\sigma^-_u\right)
\ee
with the new components
\begin{eqnarray}
S_z(\theta_L) & = & S_z \cos\theta_L +i \sin\theta_L \frac{S_+-S_-}{2} \nonumber \\
S_+(\theta_L) & = & \frac{S_+ + S_-}{2} +\cos \theta_L \frac{S_+-S_-}{2} +iS_z\sin\theta_L \\
S_-(\theta_L) & = & \frac{S_+ + S_-}{2} -\cos \theta_L \frac{S_+-S_-}{2} -iS_z\sin\theta_L  \nonumber.
\end{eqnarray}
In order to solve the left boundary equation we need to impose
\be
\bra{ \phi} S_-(\theta_L) = 0 , \quad \bra{\phi} S_z(\theta_L)= p \bra{\phi}
\ee
Using \eref{1-18} these conditions are satisfied if the coherent state parameter $\phi$ is chosen to be
\bel{1-20}
\phi = i \tan{(\theta_L/2)} .
\ee

In order to prove this result we point out
\eref{1-18}, \eref{1-19} can be used to express
vectors of the form $\bra{\phi} (a + b S_z + c S_+ + d S_-)$ that appear in the boundary equations
just in terms of e.g. $\bra{\phi} (a' + d' S_-)$, and similarly for
the action on ket-vectors $\ket{\psi}$.
The choice \eref{1-20} leads to
\be
S = \bra{\phi} \Omega^{\otimes {N}} \ket{\psi} = i p \sigma^z_u \otimes \tilde{S} + \sigma^+_u  \otimes  W
\ee
where $\tilde{S} = \bra{\phi}  \Omega^{\otimes {N-1}} \ket{\psi}$ and
$W = i \bra{\phi}  (S_+ + S_-) \Omega^{\otimes {N-1}} \ket{\psi}$.
Moreover,
\be
S S^\dagger= |p|^2 \mathds{1} \otimes
\tilde{S}{S}^\dagger - ip \sigma_u^- \otimes \tilde{S}W^\dagger -(ip)^* \sigma_u^+\otimes W\tilde{S}^\dagger +
\sigma_u^+ \sigma_u^- \otimes WW^\dagger.
\ee
Now, on the one hand we see that the action of the left dissipator $\mathcal{D}^L$ leads to
\be
\mathcal{D}^L(SS^\dagger)= 2\Gamma |p|^2 \sigma_u^z \otimes \tilde{S}\tilde{S}^\dagger +
\Gamma ip \sigma_u^- \otimes \tilde{S}W^\dagger +
\Gamma (ip)^* \sigma_u^+ \otimes W\tilde{S}^\dagger .
\ee
On the other hand the left contribution of the unitary part of the Lindblad equation leads to
\bea
i \comm{H}{SS^\dagger}\|_{Left}&= &
-\left(ip +(ip)^*\right) \sigma_u^z \otimes \tilde{S}\tilde{S}^\dagger \nonumber \\
&&- \sigma_u^- \otimes \left(1 - 2if^L (ip)\right)\tilde{S}W^\dagger \nonumber \\
&&- \sigma_u^+ \otimes \left(1 + 2if^L (ip)^* \right)W\tilde{S}^\dagger  .
\eea
Comparing the two contributions gives the solution for the representation parameter
\bel{1-21}
p= \frac{i}{\Gamma-2if^L} .
\ee

The right boundary is treated along the same lines.
The right-end state is polarized in the  $(x,z)$ plane in a direction $v$ generated by the rotation
\be
U=e^{i\frac{\theta_R}{2} \sigma^y}\; ,
\ee
where we take as reference the $-z$-direction. Going through similar steps as above we
impose the cancellation
\be
S_+(\theta_R) |\psi\rangle = 0.
\ee
With \eref{1-19} this yields to
\bel{1-22}
\psi=- \tan{(\theta_R/2)}.
\ee
In order to fulfill the stationarity condition \eref{stationary}, together with \eref{1-19} one needs to
impose $f^R=-f^L$ such that the representation parameter takes the value given in \eref{1-21}.
Interestingly, this condition turns out to allow for the inclusion of
a Dzyaloshinsky-Moriya
interaction in the XXZ-Hamiltonian \cite{KPSlong} which is the key ingredient in
the Lagrange-multiplier approach of \cite{Anta98} to current-carrying states of
quantum spin systems.

In conclusion, the solution of the completely polarized twisted case with a
polarization on the left in the $(y,z)$ plane and in the right in the $(x,z)$ plane
is given by the Matrix Product Ansatz for $S$ in the form \eref{1-4} with coherent
state parameters  \eref{1-20}, \eref{1-22} and representation parameter \eref{1-21}.
At $\theta_L=\theta_R= 0$ and vanishing boundary fields $f^R=f^L=0$
one recovers the untwisted solution \cite{Pros11b}.

The model with a twist is fundamentally different from the
untwisted one, which can be seen by studying one- and two-point
functions in the steady state. Note that in the isotropic model,
all three spin projections $\sigma _{n}^{x},$ $\sigma _{n}^{y}$
and $\sigma _{n}^{z}$ are locally conserved, i.e.,
$\frac{d}{dt}\sigma _{n}^{\alpha } = \hat{\jmath}_{n-1,n}^{\alpha
} - \hat{\jmath}_{n,n+1}^{\alpha }$, where
$\hat{\jmath}_{n,n+1}^{\alpha } = 2 \sum_{\beta ,\gamma }
\varepsilon _{\alpha \beta \gamma} \sigma _{n}^{\beta}  \sigma
_{n+1}^{\gamma}$ ($\varepsilon _{\alpha \beta \gamma } $ being
Levi-Civita symbol). This leads to three different steady state
currents $j^{\alpha} = \langle \hat{\jmath}_{n,n+1}^{\alpha}
\rangle$ for $\alpha =x,y,z$. In the untwisted model ($\theta
_{L}=\theta_{R}=0 $) two out of three one-point correlations
vanish in the steady state, $\langle \sigma
_{n}^{x}\rangle=\langle \sigma _{n}^{y} \rangle=0$ for all $n$,
corresponding to trivial flat $x-$ and $y-$ magnetization profiles
along the chain. Also, two out of three spin currents are
completely suppressed in the untwisted setup  $j^{x}=j^{y}=0$. In
a model with a twist, \textit{neither} of the one-point functions
vanishes, and \textit{all three} spin currents $j^{x},j^{y},j^{z}$
are generically nonzero.

In order to see this, we note that in the usual untwisted model
with Lindblad operators being creation/annihilation operators
\cite{Pros11a,Pros11b} the steady state is invariant under a
parity symmetry $\rho=U \rho U$ where $U=(\sigma ^{z})^{\otimes
_{N}}=U^{-1}$ \cite{PopkovLivi}. Any physical observable that
changes sign under the parity operation has to vanish in the
steady state,  e.g. $\langle \sigma _{n}^{x}\rangle = \Tr \left( U
\sigma _{n}^{x} U \rho \right) = -\Tr \left( \sigma _{n}^{x}\rho
\right) = -\langle \sigma _{n}^{x}\rangle $, from which $\langle
\sigma_{n}^{x}\rangle =0$ follows. In this way one readily obtains
$\langle \sigma _{n}^{x} \rangle = \langle \sigma _{n}^{y} \rangle
= j^{x} = j^{y} = 0$.

In the isotropic model with a twist the parity symmetry is broken,
but its place is taken by another symmetry, which we specify here
for twisting angles $\theta _{R}= - \theta _{L}= \pi/2$: It
involves left-right reflection $R(A\otimes B\otimes ...\otimes
C)=(C\otimes ....\otimes B\otimes A)R$, global rotation in the
$XY$-plane $U_{rot}=diag(1,i)^{\otimes _{N}}$ and $\Sigma
_{x}=(\sigma^{x})^{\otimes _{N}}$ and reads $\rho =V \rho
V^{\dagger }$, where $V=\Sigma _{x} U_{rot} R$
\cite{SlavaXYtwist}. It is straightforward to check that neither
of the set of observables $\langle \sigma _{n}^{\alpha }\rangle
,j^{\alpha }$,  changes sign under the $V$ symmetry, and therefore
they are generically nonzero. The symmetry $V$ does, however, give
rise to nontrivial relations between the observables, e.g.
$j^{x}=-j^{y} $, $\langle \sigma _{n}^{z}\rangle =- \langle
\sigma_{N-n}^{z}\rangle $, etc..

The major novelty of our approach is the exact MPA solution by a
quadratic algebra which turns out to be the symmetry algebra of
the unitary evolution of the bulk part of the Hamiltonian.
Remarkably, this MPA solves the stationary Lindblad equation even
though both the quantum boundary fields and the Lindblad
dissipators destroy this symmetry. We expect that Lindblad
equations for other open boundary-driven many-body quantum systems
with a $q$-deformed non-Abelian bulk symmetry can be solved in a
similar fashion. Since representations of the corresponding
quantum algebras are known, exact results for observables become
available. An open problem is the relationship of the MPA to the
integrability of the bulk Hamiltonian and hence to the extension
of the MPA approach to dynamical observables. Work on
eigenfunctions \cite{Stin95,Alca04} and recent exact results by
Eisler for the density matrix with bulk Lindblad terms
\cite{Eisl11} hint at this possibility.

\textbf{Acknowledgements}. V.P. thanks T. Prosen for pointing out
that the model with a $\pi/2$-twist is likely be  MPA-solvable.
D. K. acknowledges ANR for support through the project ANR-09-BLAN-0098-01.
G.M.S. thanks the University of Florence and the University of Lorraine for kind
hospitality.

\end{document}